# Water Heterostructure Photodetector for Calculation of Semiconductor Minority Carrier Lifetime

*Can Wang[1], Renyu Yang[1], Huikai Zhong[3], Mingjia Zhi[4], Shisheng Lin[1,2]\**

Shisheng Lin

College of Information Science and Electronic Engineering, Zhejiang University,

State Key Laboratory of Modern Optical Instrumentation, Zhejiang University,

Zhejiang University, 310027, Hangzhou, People's Republic of China.

Email: shishenglin@zju.edu.cn.

Can Wang, Renyu Yang

College of Information Science and Electronic Engineering, Zhejiang University,

Zhejiang University, 310027, Hangzhou, People's Republic of China.

Huikai Zhong

School of Mechanical and Electronic Engineering, East China University of Technology, Nanchang 330013, China

Mingjia Zhi

School of Materials Science and Engineering, Zhejiang University,

Institute for Composites Science Innovation (InCSI), Zhejiang University,

Zhejiang University, 310027, Hangzhou, People's Republic of China.

## Abstract

The minority carrier lifetime of semiconductor materials is a crucial performance parameter for optoelectronic devices. However, the existing minority carrier lifetime measurement techniques necessitate delicate optical measurement systems and harmful treatment of the samples, which will definitely cause great constraints on the further development of the semiconductor industry. Here, an off-junction graphene/water/silicon photodetector is realized based on the charming dynamic polarization process of water molecules at the water/silicon and water/graphene interface, which shows a typical responsivity and detectivity of 36.55 mA $W^{-1}$ and 1.62×$10^{12}$ Jones respectively under 890 nm illumination with a distance of 0.2 cm away from the junction. This pulse-like photo-response arises from the diffusion and drift current toward the water/silicon interface generated by the illumination on an off-junction spot. Furthermore, by measuring the photo-current at a different distance from the junction combined with an exponential fitting method, the minority carrier lifetime calculation of silicon can be performed with the maximum accuracy rate reaches 98.8%, where water plays the key role of deducing the carrier lifetime. This study provides a straightforward method that paves the way for future minority carrier lifetime tests of semiconductor materials, which hold great potential

**for developing the nondestructive tests(NDT) in semiconductor industry.**

# 1. Introduction

Minority carrier lifetime stands as one of the crucial parameters[1-3] for assessing the quality of semiconductor materials, which can reflect the defects and impurities in semiconductor materials. Through measuring minority carrier lifetime[4-7], the semiconductor device design and manufacturing processes can be optimized, ultimately leading to enhanced performance of devices. For optoelectronic devices like photodetectors and solar cells, the quantum efficiency of the device can be enhanced by a higher minority carrier lifetime in semiconductor materials[8-9], implying that the measurement of minority carrier lifetime determines the optimization of optoelectronic devices[10-12].

Techniques for measuring minority carrier lifetime predominantly fall into two categories: steady-state and transient methods[13-16]. Transient methods, particularly employing photoconductance decay (PCD), represent a common approach to determining minority carrier lifetime[17]. However, existing minority carrier lifetime measurement techniques still have some limitations. These techniques necessitate intricate optical structures or electrical measurement components[18] and demand stringent sample preparation, including passivation of the semiconductor material being measured[19], which will cause irreversible effects on the interface of semiconductor materials. With the continuous advancement of semiconductor devices, there is an urgent need for a novel minority carrier lifetime measurement technique to rapidly measure the minority carrier lifetime of semiconductor materials.

Photodetectors, with the ability to convert light signals into electrical signals, find widespread applications in critical fields for extracting information from the environment[20-25]. However, the optical current response of most photodetectors is based on carrier transport processes parallel to the built-in field[26-29]. Thus, neglecting carrier transport processes perpendicular to the built-in field results in the loss of crucial information. In recent years, the unique physical mechanisms at solid-liquid interfaces have come to attention[30-31], leading to a focus on liquid-based photodetectors. Among these, the semiconductor/polarized liquid/semiconductor junction structure has undergone extensive research[32-35].

Based on the polarization process of the water molecules, an off-junction photodetector was realized. Illuminating the semiconductor outside the junction position, photogenerated carriers transport to the junction position under the influence of concentration and potential gradients, triggering the water polarization process and thereby generating the photocurrent. This process involves carrier transport perpendicular to the built-in electric field within the junction area, thus compensating for the information loss in junction-based photodetectors. Furthermore, by measuring the photocurrent at different distances from the junction and fitting the current-distance curve with the exponential fitting method, the decay coefficient is calculated which can be used for calculating the minority carrier lifetime with the maximum accuracy rate reaches 98.8%. Meanwhile, the photodetector exhibits excellent performance in photodetection, which shows a typical responsivity and detectivity of 36.55 mA $W^{-1}$ and $1.62\times10^{12}$ Jones respectively under 890 nm illumination with a distance of 0.2 cm away from the junction.

## 2. Material and methods

The basic structure of the off-junction graphene/water/silicon (Gr/W/Si) photodetector is illustrated in **Figure 1**a. The polarized liquid is encapsulated between the graphene and n-type silicon, forming the detection configuration of the graphene/water/silicon junction. Under the illumination outside the junction, the current and voltage signals are recorded through the silver electrode on the graphene and the gold electrode on the backside of silicon. To ensure that the incident light does not illuminate at the Gr/W/Si junction, a light shield layer is covered at the top of the junction. The incident light is respectively irradiated on the off-junction area in the device with a light shield layer and the device is completely isolated from the light. The consistency of the current signals between the two devices indicates that the light shield layer can effectively isolate the incident light, ensuring that the light only illuminates the off-junction silicon area.

The graphene utilized in the device is transferred onto a transparent PET substrate via a wet transfer method. Figure 1b shows the Raman spectrum of graphene. The Raman spectroscopy characterization of used graphene shows the characteristic peaks located at 1590.71 and 2693.67 $cm^{-1}$, corresponding to G and 2D peaks, individually. Around the D peak (about 1350 $cm^{-1}$), the intensity is pretty weak, which indicates few defects in graphene. Moreover, the intensity of the 2D peak is significantly greater than that of the G peak, consistent with the properties of monolayer graphene. Under the light switching, the current-time curve of the device under zero bias is shown in Figure 1c. Illuminating by the light source with the wavelength of 890 nm at the distance of 0.2 cm from the junction, the pulse-like current reached 2.12 μA. The I-V curve of the device in the dark state and under 890 nm illumination is shown in Figure 1d, which shows the rectification characteristic of the device. The operation of graphene/water/silicon photodetectors primarily relies on transient photocurrents, with no distinct steady-state photocurrent observed under continuous illumination.

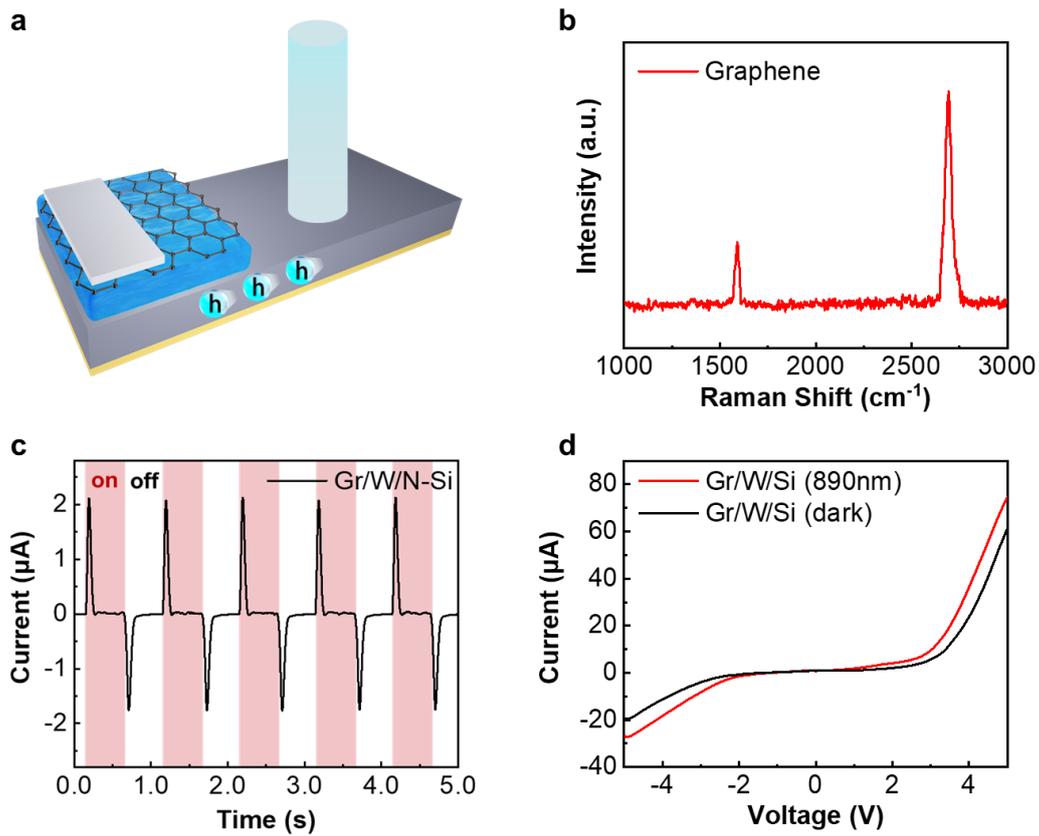

**Figure 1.** a) Schematic diagrams of the off-junction photodetector. b) Raman spectrum of graphene. c) Time-resolved current of the device under 890 nm illumination under zero bias. d) I-V curve of the device under dark and illumination of 890 nm light.

## 3. Theory/calculation

The working mechanism of the device under light switching is illustrated in **Figure 2**. As graphene, n-type silicon, and deionized water come into initial contact, holes, and electrons are accumulated on the surface of graphene and n-type silicon respectively due to the difference in Fermi level among the three materials. The accumulation of surface carriers on graphene and silicon leads to the organized alignment of polarized water molecules, where oxygen atoms orient towards the n-type silicon and hydrogen atoms orient towards the graphene, thus shielding the surface charges on both graphene and n-type silicon. This state is referred to as the initial state, as shown in Figure 2a.

When the silicon surface is illuminated by the light source, the holes generated by the light accumulate at the silicon surface. The hole concentration at the illumination position is higher compared to the non-illuminated position. Meanwhile, a positive potential in n-type silicon is generated when the silicon is illuminated. In contrast, at the Gr/W/Si position, the energy bands bend upwards, leading to a negative potential. As a result, driven by both concentration gradient and potential gradient, the holes transport to the silicon surface of the Gr/W/Si junction, as shown in Figure 2b. A significant number of holes accumulate at the water/silicon interface, leading to the polarization of numerous water molecules and consequently generating a positive current.

Under continuous illumination, the transport and recombination of holes at the water/silicon interface attain dynamic equilibrium, as shown in Figure 2c, resulting in no photocurrent generation. Upon light source cessation, a multitude of minority carriers at the water/silicon interface recombine, as shown in Figure 2d. Numerous water molecules depolarize due to the reduction of minority carriers at the water/silicon interface, thereby generating a negative current. When minority carriers recombine completely, the device returns to the initial state.

The doping concentration of the silicon used in the device can be calculated using the following formula:

$$\rho_n = \frac{1}{q \times N_D \times \mu_n} \tag{1}$$

where $\rho$ is the resistivity of the n-type silicon, $q$ is the charge constant, $N_D$ is the carrier concentration, $\mu_n$ is the intrinsic electron mobility, which is 1450cm² V⁻¹ s⁻¹. Once the doping concentration is calculated, the Fermi level of the silicon can be further computed by the following formula:

$$E_{F-n} = E_i + k_B \times T \times \ln\left(\frac{N_D - N_A}{n_i}\right) \quad (2)$$

where $E_{F-n}$ is the Fermi level of the n-type silicon, $E_i$ is the intrinsic Fermi level, $k_B$ is the Boltzmann constant, $T$ is the temperature, $n_i$ is the electron concentration.

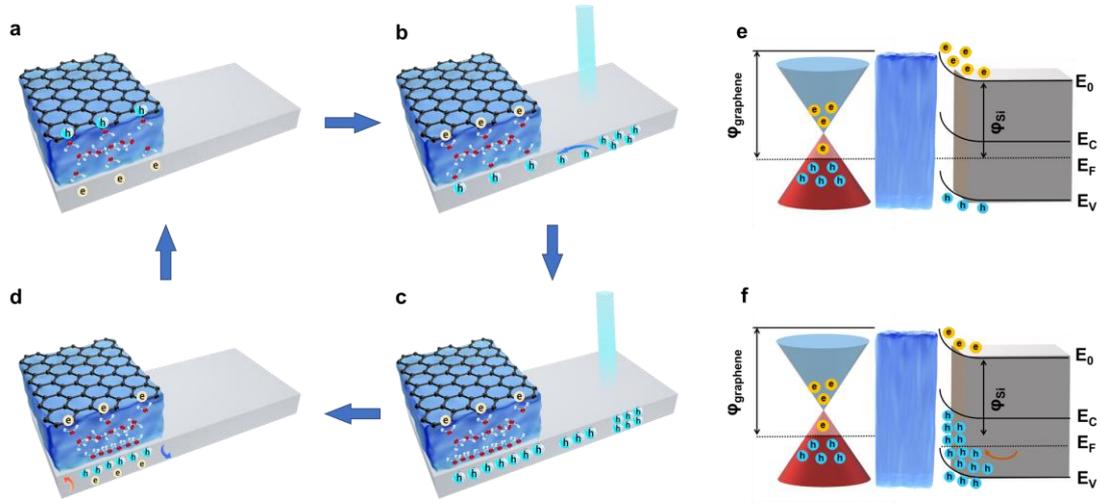

**Figure 2.** a) The initial state of the off-junction Gr/W/Si photodetector after contact. b) The polarization state caused by the potential and concentration gradient. c) The saturation state due to the dynamic equilibrium. d) The depolarization state caused by the recombination process. e) The band structure of the Gr/W/Si junction after contact. f) The band structure of the Gr/W/Si junction under the illumination.

The Fermi level of the n-type silicon with a resistivity of 10 Ω cm is calculated to be 4.32 eV by Formula 1 and 2. The work function of graphene is calculated to be 4.84 eV according to the Raman Spectrum and the chemical potential of water is 4.26 eV. Due to the differences in work function, the energy band bends when graphene, water, and silicon come into contact. The electrons in n-type silicon diffuse to the water/silicon interface and the holes in graphene diffuse to the water/graphene interface. The detailed energy band structure of the Gr/W/Si junction is shown in Figure 2e.

The band structure of the Gr/W/Si junction under the illumination is shown in Figure

2f. Illuminating with the wavelength of 890 nm outside the junction, the photogenerated holes on the n-type silicon diffuse from the illumination position to the junction position. As holes continue to diffuse to the junction position, the electrons on the water/silicon interface are recombined leaving behind a significant number of holes. Due to variations in charge polarity and charge concentration at the water/silicon interface, water molecules are polarized to shield the change of charge amount on the silicon surface, thereby generating a photogenerated current.

The difference between the Fermi level and the polarity of the interface charge determines the performance of the device. Conduct a photocurrent response experiment by replacing n-type silicon with p-type silicon with a resistivity of 10 Ω cm in the device. Compared to the Fermi level difference of 0.54eV between n-type silicon and graphene, the difference of Fermi level between p-type silicon only reaches 0.06eV, resulting in the significantly smaller photocurrent of the graphene/water/p-type silicon photodetector than that of the graphene/water/n-type photodetector. Additionally, majority carriers in n-type silicon are electrons, while majority carriers in p-type silicon are holes. Under illumination outside the junction, n-type silicon excites holes, causing the oxygen atoms of water molecules to polarize towards the n-type silicon, thereby generating forward photocurrent. Conversely, p-type silicon excites electrons, causing the hydrogen atoms of water molecules to polarize towards the p-type silicon, resulting in reverse photocurrent.

Due to the illumination outside the junction, detection can be performed by varying the distance of the light source. In this process, the photocurrent involves carrier transport perpendicular to the built-in field, introducing novel physical implications. As the distance between the light source and the junction position changes, the distribution of carriers on the semiconductor surface varies, as shown in **Figure 3**a. To further explore the application of the device, the light detection performance of the light source at a certain distance should be tested first. With the increase in optical power, a larger number of charge carriers are excited in the off-junction region. Consequently, more carriers transport towards the junction position, leading to the polarization of water molecules and resulting in a larger photocurrent.

The $T_{rise}$ and $T_{fall}$ were calculated to be 26 ms and 37 ms respectively. The responsivity can be calculated by the following:

$$R = \frac{I_l - I_d}{Ps} \tag{3}$$

where $I_l$ is the current under illumination, $I_d$ is the current under dark, $P$ is the incident light power density, $s$ is the effective illumination area. Meanwhile, the detectivity can be calculated by the following:

$$D^* = \frac{\sqrt{s} \times R}{\sqrt{2 \times q \times I_d}} \tag{4}$$

where $q$ is the charge constant. The maximum R is calculated to be 36.55 mA W$^{-1}$ under 890 nm illumination by Formula 3. According to the semiconductor intrinsic absorption theory, the energy of incident photons needs to satisfy the following equation:

$$E = hv = \frac{hc}{\lambda} \geq E_g \tag{5}$$

where $h$ is the Planck constant, $v$ is the frequency of light, $c$ is the speed of light, $\lambda$ is the wavelength of light, and $E_g$ is the bandgap width of the semiconductor.

For silicon, the bandgap width is 1.12 eV, and the absorption edge wavelength of silicon is calculated to be 1107 nm by Formula 5. When the incident wavelength exceeds 1000 nm, the responsivity of the device rapidly decreases, consistent with the semiconductor intrinsic absorption theory. The device achieves peak responsivity at a wavelength of 890 nm, consistent with the wavelength-dependent characteristics of the graphene/n-type silicon junction photodetector. Figure 3b shows the $R$ and $D^*$ of the transient photocurrent under different incident light power. The maximum $R$ and $D^*$ were calculated as 36.55 mA W$^{-1}$ and 1.62×10$^{12}$ Jones by Formula 3 and 4 respectively. The responsivity and detectivity gradually decrease as the optical power grow indicating that more photo-generated carriers may be captured by defect states as the incident optical power density increases. Only a portion of the photo-generated carriers can participate in the generation of photocurrent, leading to a decrease in responsivity and detection rate with increasing incident optical power. Simultaneously, according to Formula 3 and 4, the saturation of the photocurrent and the increasing optical power will lead to a decrease in responsivity and detectivity.

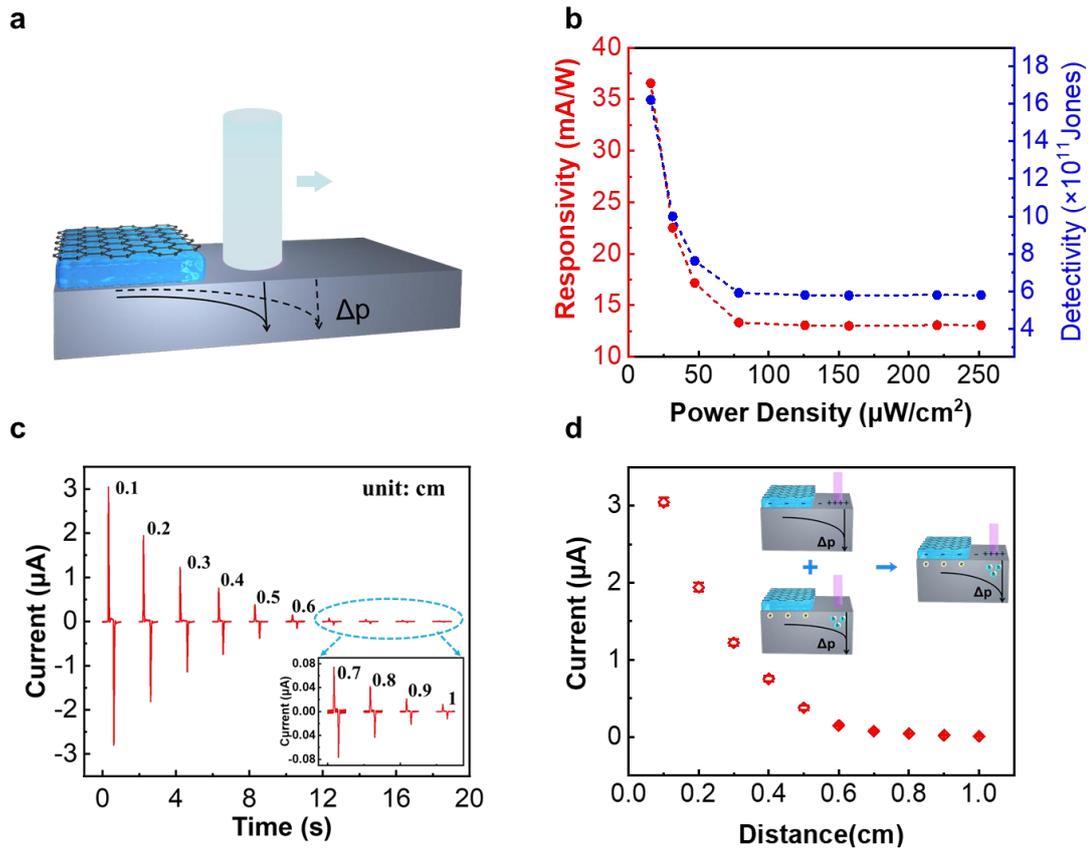

**Figure 3.** a) The schematic diagram of minority carrier concentration distribution with distance variation. b) The relationship between responsivity/detectivity and the incident light power density under 890nm illumination. c) The current-time curve with distance variation. d) Distance dependence of the light-switching-induced transient current. The insert is the schematic diagram of carrier distribution under the coupling of drift and diffusion process.

Illuminating the silicon outside the junction with the switching light at the wavelength of 532 nm results in the optical response, as shown in Figure 3c. As the distance increases, the transient photocurrent diminishes. When the light source is sufficiently far away, the photocurrent will attenuate to zero. Plotting the corresponding transient photocurrents against distances, Figure 3d is obtained. With increasing distance, the transient photocurrent exhibits exponential decay which is similar to the decay of minority carrier concentration in semiconductors, which indicates the photocurrent contains information about the minority carrier lifetime due to the generation of photocurrent involving the carrier transport process. Under illumination outside the junction, both the potential gradient and carrier concentration gradient are generated, leading to the drift and diffusion processes of carriers. Thus, the total current generated by the device is the coupling of drift current and diffusion current.

Incident photons are absorbed by the semiconductor, leading to the excitation of numerous minority carriers and thus establishing the concentration gradient of minority carriers from the illumination position to the junction position. Driven by the concentration gradients, numerous minority carriers diffuse to the junction position, triggering the change in the carrier concentration at the water/silicon interface and thereby generating a transient photocurrent. According to semiconductor physics, the continuity equation describing the distribution of minority carrier concentration is as follows.

$$\frac{\partial p}{\partial t} = D_p \frac{\partial^2 p}{\partial x^2} - \mu_p \varepsilon \frac{\partial p}{\partial x} - \frac{\Delta p}{\tau} \tag{6}$$

where $p$ is the hole concentration, $D_p$ is the diffusion coefficient, $\mu_p$ is the hole mobility, $\tau$ is the lifetime of the hole, $\varepsilon$ is the electric field, $x$ is the distance and $t$ is the time. The surface saturation minority carrier density at the water/silicon interface decreases when the light source moves away from the junction position due to the loss of minority carriers during the transport process as the transport distance increases. Consequently, the number of polarized water molecules at the water/silicon interface decreased, leading to a decrease in the transient photocurrent. The exponential decrease trend of transient photocurrent with distance variation conforms to the continuity equation. Due to the loss of minority carriers in the diffusion process, the distribution of surface carriers on the semiconductor follows the formula.

$$\Delta p = (\Delta p)_0 e^{-\frac{x}{L_p}} \tag{7}$$

where $\Delta p$ represents the non-equilibrium minority carrier concentration, $(\Delta p)_0$ represents the initial non-equilibrium minority carrier concentration under illumination, x stands for the minority carrier travel distance, and $L_p$ represents the minority carrier diffusion length, with $L_p = \sqrt{D_p \tau_p}$. Based on the theory of liquid polarization, the variation in the carrier concentration at the semiconductor surface induces the polarization and depolarization processes of water molecules. This process determines the transient current of the liquid-based photodetector implying that the photocurrent is related to the semiconductor surface carrier concentration as follows:

$$I = \alpha \cdot \Delta p \tag{8}$$

where $I$ represent the magnitude of the photocurrent, and $\alpha$ is the current conversion coefficient, representing the relationship between the carrier concentration and the photocurrent, which should be a constant. The photocurrent reaches the maximum value when the light source directly illuminates at the junction. With the increasing distance between the light source and the junction, the photocurrent gradually decreases due to the limitation of the carrier lifetime. The fitting curves of the current and distance exhibit exponential decay, consistent with the decay form of the minority carrier concentration. According to Formula 8, the conversion between the carrier concentration and the current can be performed. Since the light source illuminates the semiconductor surface, both the concentration gradient and potential gradient collectively drive the minority carriers from the illumination position to the junction position. Therefore, it is necessary to consider both diffusion and drift movements. Thus, the distance of carrier transport should be the sum of the drift distance and the diffusion distance, deriving the following fitting formula:

$$I = I_s e^{-\beta d} \tag{9}$$

where $\beta$ is the decay coefficient, representing the decay of carrier concentration during the transport process under the driving forces of concentration and potential gradients. The decay coefficient can be obtained by fitting the current-distance curve.

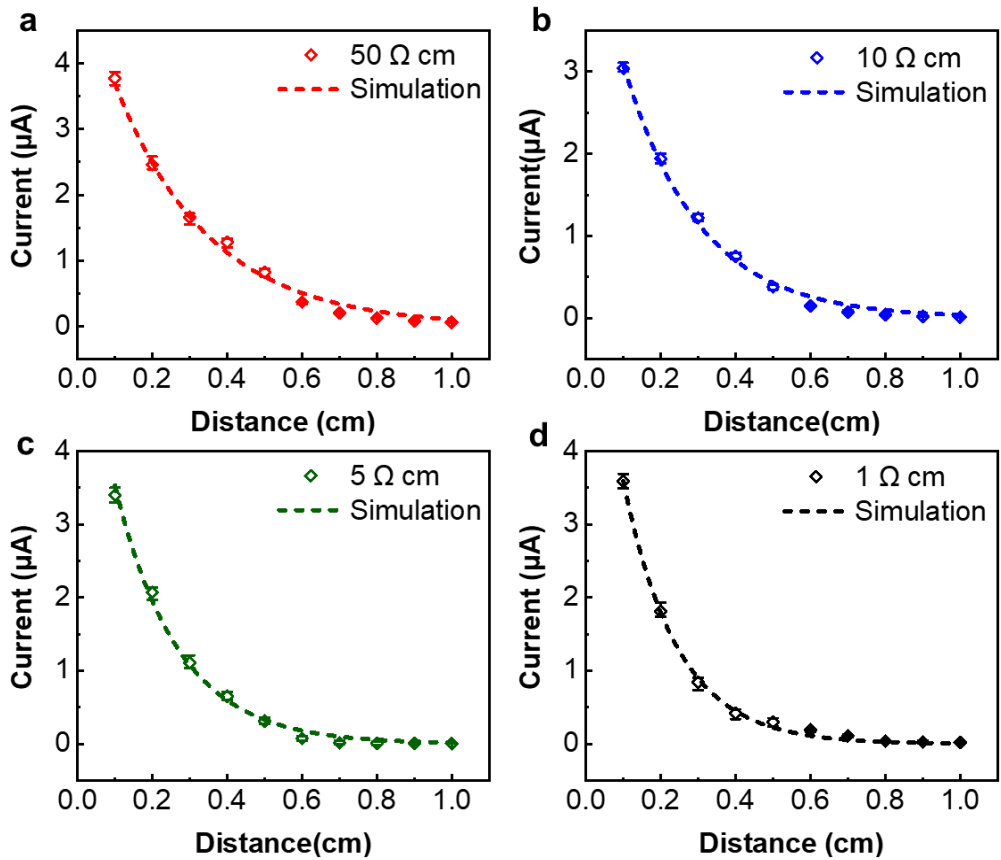

**Figure 4.** Current-distance curves of silicon with different resistivities, a) 50 Ω cm. b) 10 Ω cm. c) 5 Ω cm. d) 1 Ω cm.

Silicon with different resistivities possess distinct doping concentrations, leading to the difference in minority carrier lifetime. Thus, experiments related to minority carrier lifetimes can be conducted by replacing silicon wafers with different resistivities. Therefore, extracting information related to the minority carrier lifetime from transient photocurrents of the devices can be realized. By altering the distance between the light source and the junction, the current-distance curves were measured for silicon with resistivities of 50 Ω cm, 10 Ω cm, 5 Ω cm, and 1 Ω cm, as depicted in **Figure 4**. Utilizing Formula 9 to fitting the curves, the expression for the current-distance relationship for silicon with different resistivities is as follows: $I = 5.5 \cdot e^{-3.976d}$ for 50 Ω cm, $I = 5.06 \cdot e^{-4.944d}$ for 10 Ω cm, $I = 6.4 \cdot e^{-5.94d}$ for 5 Ω cm, and $I = 7.2 \cdot e^{-6.947d}$ for 1 Ω cm. The decay coefficients are determined to be 3.976, 4.944, 5.940, and 6.947, representing the decay of photocurrent with distance and reflecting information about minority carrier lifetime. A larger decay coefficient indicates faster decay of photocurrent, which implies a shorter minority carrier lifetime.

**Table 1. Summary of experiment results**

| Resistivity (Ω cm) | Theoretical minority carrier lifetime (s) | Decay coefficient β | Fitting minority carrier lifetime (s) |
|---|---|---|---|
| 50 | $5 \times 10^{-4}$ | 3.976 | $4.94 \times 10^{-4}$ |
| 10 | $3 \times 10^{-4}$ | 4.944 | $3.19 \times 10^{-4}$ |
| 5 | $2 \times 10^{-4}$ | 5.940 | $2.21 \times 10^{-4}$ |
| 1 | $1.5 \times 10^{-4}$ | 6.947 | $1.61 \times 10^{-4}$ |

Further decomposition of the decay coefficient and substitution of known quantities can deduce the minority carrier lifetime. Subsequently, the minority carrier lifetimes for the silicon with resistivities of 50 Ω cm, 10 Ω cm, 5 Ω cm, and 1 Ω cm were calculated as $4.94 \times 10^{-4}$ s, $3.19 \times 10^{-4}$ s, $2.21 \times 10^{-4}$ s, and $1.61 \times 10^{-4}$ s respectively, as indicated in **Table 1**. Compared to the theoretical data, the fitted data has an error of less than 10.5%, as the minimum error is 1.2%. Thus, we have developed a method for detecting the minority carrier lifetime with an accuracy rate as high as 98.8%. The fitting minority carrier lifetimes closely align with the theoretical values, affirming the capability of this method in measuring minority carrier lifetimes.

## 4. Conclusion

In summary, we developed an off-junction water photodetector that transcends the limitation of PN junction-based photodetectors by measuring the transport process of minority carriers. Under illumination, the potential and concentration gradients drive the accumulation of minority carriers at the water/silicon interface, inducing the polarization and the depolarization process of water molecules and consequently resulting in transient photocurrent. By varying the position of the light source outside the junction, the minority carrier lifetime of semiconductors can be calculated by fitting the current by the exponential formula. Compared to the theoretical value, the calculated minority carrier lifetime has an error of less than 10.5%, as the accuracy rate reaches 98.8%. Besides, this device exhibits excellent optoelectronic detection capabilities, whose responsivity and detectivity under 890 nm illumination are 36.55 mA W$^{-1}$ and $1.62\times10^{12}$ Jones.

## Author Contributions

S. Lin designed the experiments, analyzed the data, conceived the study, and wrote the paper. C. Wang and R. Yang carried out the experiments, discussed the results, and wrote the paper. H. Zhong and M. Zhi discussed the results and assisted with experiments. All authors contributed to the preparation of the manuscript.

## Acknowledgements

S. S. Lin thanks the support from the National Natural Science Foundation of China (Grant Nos. 51202216, 51551203, 61774135 and 62474161), Special Foundation of Young Professor of Zhejiang University (Grant No. 2013QNA5007), and Outstanding Youth Fund of Zhejiang Natural Science Foundation of China (Grant No. LR21F040001).

## Conflict of Interest

The authors declare no conflict of interest.

## Data Availability Statement

The data that support the findings of this study are available from the corresponding author upon reasonable request.

## Keywords

graphene, photodetector, off-junction detection, minority carrier lifetime.

## Uncategorized References

[1] Y. Jung, A. Vacic, D. E. Perea, S. T. Picraux, M. A. Reed, *Advanced Materials* **2011**, 23, 4306.

[2] G. T. Reed, G. Mashanovich, F. Y. Gardes, D. J. Thomson, *Nature Photonics* **2010**, 4, 518.

[3] P. Uprety, I. Subedi, M. M. Junda, R. W. Collins, N. J. Podraza, *Scientific Reports* **2019**, 9, 19015.

[4] C.-H. Chu, M.-H. Mao, C.-W. Yang, H.-H. Lin, *Scientific Reports* **2019**, 9, 9426.

[5] S. Dev, Y. Wang, K. Kim, M. Zamiri, C. Kadlec, M. Goldflam, S. Hawkins, E. Shaner, J. Kim, S. Krishna, M. Allen, J. Allen, E. Tutuc, D. Wasserman, *Nature Communications* **2019**, 10, 1625.

[6] H. Hempel, C. J. Hages, R. Eichberger, I. Repins, T. Unold, *Scientific Reports* **2018**, 8, 14476.

[7] A. Baumann, J. Lorrmann, D. Rauh, C. Deibel, V. Dyakonov, *Advanced Materials* **2012**, 24, 4381.

[8] J. Guo, B. Wang, D. Lu, T. Wang, T. Liu, R. Wang, X. Dong, T. Zhou, N. Zheng, Q. Fu, Z. Xie, X. Wan, G. Xing, Y. Chen, Y. Liu, *Advanced Materials* **2023**, 35, 2212126.

[9] X. Yang, Y. Fu, R. Su, Y. Zheng, Y. Zhang, W. Yang, M. Yu, P. Chen, Y. Wang, J. Wu, D. Luo, Y. Tu, L. Zhao, Q. Gong, R. Zhu, *Advanced Materials* **2020**, 32, 2002585.

[10] A. Augusto, J. Karas, P. Balaji, S. G. Bowden, R. R. King, *Journal of Materials Chemistry A* **2020**, 8, 16599.

[11] J. Geist, *Journal of Applied Physics* **1980**, 51, 3993.

[12] L. Zhan, S. Yin, Y. Li, S. Li, T. Chen, R. Sun, J. Min, G. Zhou, H. Zhu, Y. Chen, J. Fang, C.-Q. Ma, X. Xia, X. Lu, H. Qiu, W. Fu, H. Chen, *Advanced Materials* **2022**, 34, 2206269.

[13] A. Cuevas, D. Macdonald, *Solar Energy* **2004**, 76, 255.

[14] S. Parola, M. Daanoune, A. Focsa, B. Semmache, E. Picard, A. Kaminski-Cachopo, M. Lemiti, D. Blanc-Pélissier, *Energy Procedia* **2014**, 55, 121.

[15] T. Sameshima, T. Nagao, S. Yoshidomi, K. Kogure, M. Hasumi, *Japanese Journal of Applied Physics* **2011**, 50, 03CA02.

[16] Z. Zhao, C. Xieyu, T. Zhen, *Infrared and Laser Engineering* **2019**, 48, 919003.

[17] P. J. Drummond, D. Bhatia, A. Kshirsagar, S. Ramani, J. Ruzyllo, *Thin Solid Films* **2011**, 519, 7621.

[18] L. Sirleto, A. Irace, G. F. Vitale, L. Zeni, A. Cutolo, *Optics and Lasers in Engineering* **2002**, 38, 461.

[19] S. Johnston, K. Zaunbrecher, R. Ahrenkiel, D. Kuciauskas, D. Albin, W. Metzger, *IEEE Journal of Photovoltaics* **2014**, 4, 1295.

[20] A. Pospischil, M. Humer, M. M. Furchi, D. Bachmann, R. Guider, T. Fromherz, T. Mueller, *Nature Photonics* **2013**, 7, 892.

[21] M. Engel, M. Steiner, P. Avouris, *Nano letters* **2014**, 14, 6414.

[22] Z. H. Lin, G. Cheng, Y. Yang, Y. S. Zhou, S. Lee, Z. L. Wang, *Advanced*

*Functional Materials* **2014**, 24, 2810.

[23] N. Flöry, P. Ma, Y. Salamin, A. Emboras, T. Taniguchi, K. Watanabe, J. Leuthold, L. Novotny, *Nature Nanotechnology* **2020**, 15, 118.

[24] M. Yang, Q. Han, X. Liu, J. Han, Y. Zhao, L. He, J. Gou, Z. Wu, X. Wang, J. Wang, *Advanced Functional Materials* **2020**, 30, 1909659.

[25] C. Bartolo-Perez, S. Chandiparsi, A. S. Mayet, H. Cansizoglu, Y. Gao, W. Qarony, A. AhAmed, S.-Y. Wang, S. R. Cherry, M. S. Islam, *Optics Express* **2021**, 29, 19024.

[26] B. W. H. Baugher, H. O. H. Churchill, Y. Yang, P. Jarillo-Herrero, *Nature Nanotechnology* **2014**, 9, 262.

[27] X. Feng, X. Zhao, L. Yang, M. Li, F. Qie, J. Guo, Y. Zhang, T. Li, W. Yuan, Y. Yan, *Nature Communications* **2018**, 9, 3750.

[28] J. F. Tasker, J. Frazer, G. Ferranti, E. J. Allen, L. F. Brunel, S. Tanzilli, V. D'Auria, J. C. F. Matthews, *Nature Photonics* **2021**, 15, 11.

[29] J. Chen, X. Liu, Z. Li, F. Cao, X. Lu, X. Fang, *Advanced Functional Materials* **2022**, 32, 2201066.

[30] Y. Lu, Y. Yan, X. Yu, X. Zhou, S. Feng, C. Xu, H. Zheng, Z. Yang, L. Li, K. Liu, S. Lin, *Research* **2021**, 2021.

[31] Y. Yan, X. Zhou, S. Feng, Y. Lu, J. Qian, P. Zhang, X. Yu, Y. Zheng, F. Wang, K. Liu, S. Lin, *The Journal of Physical Chemistry C* **2021**, 125, 14180.

[32] J. Li, Y. Long, Z. Hu, J. Niu, T. Xu, M. Yu, B. Li, X. Li, J. Zhou, Y. Liu, *Nature Communications* **2021**, 12, 4998.

[33] C. Liu, Y. Lu, Y. Zhang, X. Yu, C. Wang, R. Shen, S. Lin, *Cell Reports Physical Science* **2022**, 3, 101192.

[34] Y. Yan, Y. Lu, C. Liu, X. Yu, C. Wang, R. Yang, S. Lin, *Solar RRL* **2022**, 6, 2200782.

[35] Z. Qian, M. Yang, S. Lin, *Advanced Functional Materials* **2025**, 35.